\documentclass[dvipsnames,12pt]{article}
\usepackage{adjustbox}
\usepackage{amssymb}
\usepackage{amsthm}
\usepackage{amsmath}
\usepackage{authblk}
\usepackage{blkarray}
\usepackage{booktabs}
\usepackage{bm}
\usepackage{calc}
\usepackage{caption}
\usepackage{cases}
\usepackage{centernot}
\usepackage[utf8]{inputenc}
\usepackage{dcolumn}
\usepackage{float}
\usepackage[margin = 1in]{geometry}
\usepackage{graphicx}
\usepackage{lineno}
\usepackage{listings}
\usepackage{makecell}
\usepackage{mathtools}
\usepackage[authoryear,sort]{natbib}
\usepackage{pifont}

\usepackage{setspace}
\usepackage{soul}
\usepackage{subcaption}
\usepackage{tabularx}
\usepackage{tgpagella}
\usepackage{verbatim}
\usepackage{xcolor}
\usepackage{xfrac}
\usepackage[style=english]{csquotes}

\title{Phage-antibiotic therapy under density dependent bacterial defenses}
\author{Rohan Shirur\thanks{rs23bs@fsu.edu} \, and Bryce Morsky\thanks{bmorsky@fsu.edu}}
\affil{Department of Mathematics, Florida State University, Tallahassee, FL, USA}
\date{\today}

\begin{document}

\maketitle

\begin{abstract}
Phage therapy is an alternative treatment method for bacterial infections. It has shown particular promise in reducing bacterial load while preventing antibiotic resistance. Here, we develop a mathematical model of a bacterial infection within a host to study phage therapy. It incorporates interactions between phages, bacteria, the immune system, and antibiotics. Additionally, the model includes bacterial social dynamics that provide protection from treatments and the innate immune response. We analytically and numerically identify all of the equilibria of the model and derive insights regarding the overall effectiveness of phage therapy. Without phage therapy, the model exhibits bistability: bacteria populations above a threshold grow and become entrenched, while those below it can be effectively suppressed by the immune system. We find that that phages destabilize the former equilibrium, and thus in combination with the immune system are able to suppress the bacteria. We conducted bifurcation analyses, which show that the equilibrium with a suppressed population of bacteria can become unstable. In this scenario, the system undergoes oscillations. However, these oscillations --- which can be exacerbated by social dynamics --- lead to minuscule bacterial populations, and thus, in practice, phage therapy is widely effective across the parameter space. We also demonstrate how suppression can be further improved by the addition of periodic dosing of antibiotics in a combination therapy.
\end{abstract}
{\textbf{Keywords:}} antibiotic resistance, combination therapy, immune response, phage therapy, social dynamics

\section{Introduction}

The healthcare system relies heavily upon antibiotics to combat bacterial infection. However, the overreliance on and misuse of antibiotics have resulted in the emergence of antibiotic resistant strains of bacteria \citep{barbosa2000impact} with a continuing evolution of novel antibiotic resistant strains over time \citep{normark2002evolution}. In addition to mutation, antibiotic resistance can be spread rapidly across various bacterial organisms and even different bacterial species by means of horizontal gene transfer \citep{burmeister2015horizontal}. Thus, alternative treatment options are paramount to sustainably fight bacterial infections. One such treatment is phage therapy, which utilizes bacteriophages, also known as phages, to selectively eliminate bacterial pathogens \citep{gordillo2019phage,fujiki2023phage}. It has shown significant promise in fighting bacterial infections while minimizing antibiotic resistance. This is due in part to the inherent strain specificity of phages which limits indiscriminate bacterial killing \citep{zalewska2025insights}. Trials have even indicated that, unlike antibiotics, phages can leave the gut microbiome (a vital component of the immune system) largely intact when fighting bacterial infection \citep{fujiki2023phage}. Phage therapy has been demonstrated to effectively reduce bacterial load while being well-tolerated by the host with limited side effects \citep{uyttebroek2022safety}. It has also achieved therapeutic success at sub-inhibitory concentrations of antibiotics \citep{rodriguez2020quantitative}. Phage therapy can be implemented along with antibiotics in combination and adaptive therapies, the latter of which incorporate a dynamic treatment approach that adjusts relative to bacterial evolution \citep{levin2004population}. Such therapies are primarily done by carefully selecting appropriate phages and drugs, as well as their respective doses, in an attempt to eliminate the pathogenic bacteria within the host while minimizing the proliferation of antibiotic resistance.

Although such therapies are crucial in controlling infection, they are typically complimentary to the host's innate immune system, which is the primary mechanism by which bacterial infection is contained and suppressed. The immune response consists of a multifaceted response to neutralize the invading bacteria that adapts and scales based upon the type as well as the quantity of bacteria present \citep{giamarellos2012immune}. The immune system's initial defense mechanism is to deploy white blood cells, specifically neutrophils and macrophages. In the event of a more prolonged infection where the pathogenic bacteria is not immediately suppressed, two other types of white blood cells, T cells and B cells, are deployed by the immune system \citep{shepherd2020t}. T cells kill infected cells and activate other immune responses that can provide long term immunity \citep{blanden1974t}. Long term immunity is provided by the production of memory T cells, which are able to efficiently identify and respond to previously encountered pathogens \citep{blanden1974t}. In conjunction with T cells, the innate immune response also deploys B cells, which function to produce antibodies and establish long term adaptive immunity \citep{akkaya2020b}. These B cells are activated around the same time as T cells during the immune response and often require T-helper cell signaling \citep{akkaya2020b}. The primary goal of the immune system when fighting bacterial infection is to not just eradicate harmful bacteria but to instead foster long-term immunity processes that will ensure that the body will be able to adapt and respond to future infections.

In addition to the effects from treatments and the immune system, the overall outcome of a bacterial infection is linked to the behavior of the pathogenic bacteria \citep{javaudin2021intestinal}. Bacteria are social creatures, engaging in cooperative behaviors to coordinate actions to enhance their fitness, increase their chances of survival, and compete for limited resources \citep{west2007social}. They exhibit a wide array of social behaviors including, but not limited to, communication, cooperation, and group motility. These cooperative behaviors can result in rapid bacterial growth, increased virulence, and an overall higher bacteria survival rate \citep{harrison2006cooperation}. Bacterial interactions are facilitated by close proximity of bacteria \citep{ibberson2020social}. Bacteria use a communication method known as quorum sensing to transmit information to one another through the use of autoinducers and thus interact socially \citep{zhao2018production,abisado2018bacterial}. Quorum sensing can promote biofilm formation, which can prolong their survival and undermine immune responses \citep{kostakioti2013bacterial,gestel2015division}. Within biofilms, bacteria use chemical and mechanical signals to establish division of labor, resource sharing, and resistance to external threats. These processes can protect bacteria within a host from harmful conditions \citep{jefferson2004drives}. For instance, bacteria within biofilms can benefit from a $100$ to $1000$ fold increase in antibiotic resistance relative to the resistance of free swimming bacteria \citep{jarrett2015modelling}. Although these social interactions tend to benefit bacteria, they can also harm them. Current literature suggests that there exists a ``goldilocks'' zone for bacteria within a system in which they experience maximal social benefit from one another \citep{ross2009density}. At bacterial populations below the optimal range, there are simply not enough bacteria for the colony to truly reap the benefits from social behaviors \citep{west2007social}. Bacteria at populations above the optimal range experience significantly more competition for resources, and social interactions at this population level are counterproductive to each cell's best interests.

In order to effectively understand bacterial infection, phage and combination therapies, and the evolution of antibiotic resistance, it is important that we observe the mechanisms behind the interactions between therapies, the innate immune responses of various hosts, and the behavioral tendencies of bacteria. This process can be observed and analyzed through mathematical modeling techniques. This approach has shown that phage therapy, accounting for the synergistic relationship between phages and the innate immune response, can be effective in protecting normal mice from bacterial infection, but ineffective in neutropenic mice \citep{leung2017modeling}. It has also demonstrated that the variability in the outcomes of bacterial infection are most closely tied to variability in the bacterial growth rate, strength of the innate immune response, and the activation rate of the immune response \citep{barber2021predicting}. These systems can be bistable in that infections can lead to the complete elimination of the pathogenic bacteria by the immune system or uncontrolled bacterial growth. In such scenarios, the timing of the immune response is the primary factor determining outcomes, since the treatment works by ``buying more time'' for the immune system to eradicate the infection and prevent uncontrolled bacterial growth \citep{flores2022mathematical}.

Bacterial social dynamics, particularly the formation of biofilms and quorum sensing, are a critical component of the outcomes of phage therapy. Biofilms can provide protection from phages \citep{hansen2019big,singh2022good}. For example, diffusion of phages into and within the biofilm can be hindered by extracellular polymeric substances. Furthermore, phage resistant bacteria can protect phage sensitive bacteria \citep{simmons2020biofilm}, and adsorption can be ineffective due to the close proximity of bacteria \citep{eriksen2018growing}. Bacteria can also be protected by employing quorum sensing to reduce phage receptors \citep{hoyland2017quorum}. In order to understand the impacts of these defensive capabilities on the success of phage therapy, we developed and analyze a mathematical model that integrates the dynamics between phages, bacteria, antibiotics, and the innate immune response. Our primary focus is on exploring the dynamics of phage therapy and antibiotics in supplementing the innate immune response in fighting bacterial infections.

\section{Methods}

\begin{figure}[ht!]
    \centering
    \includegraphics[width=0.7\textwidth]{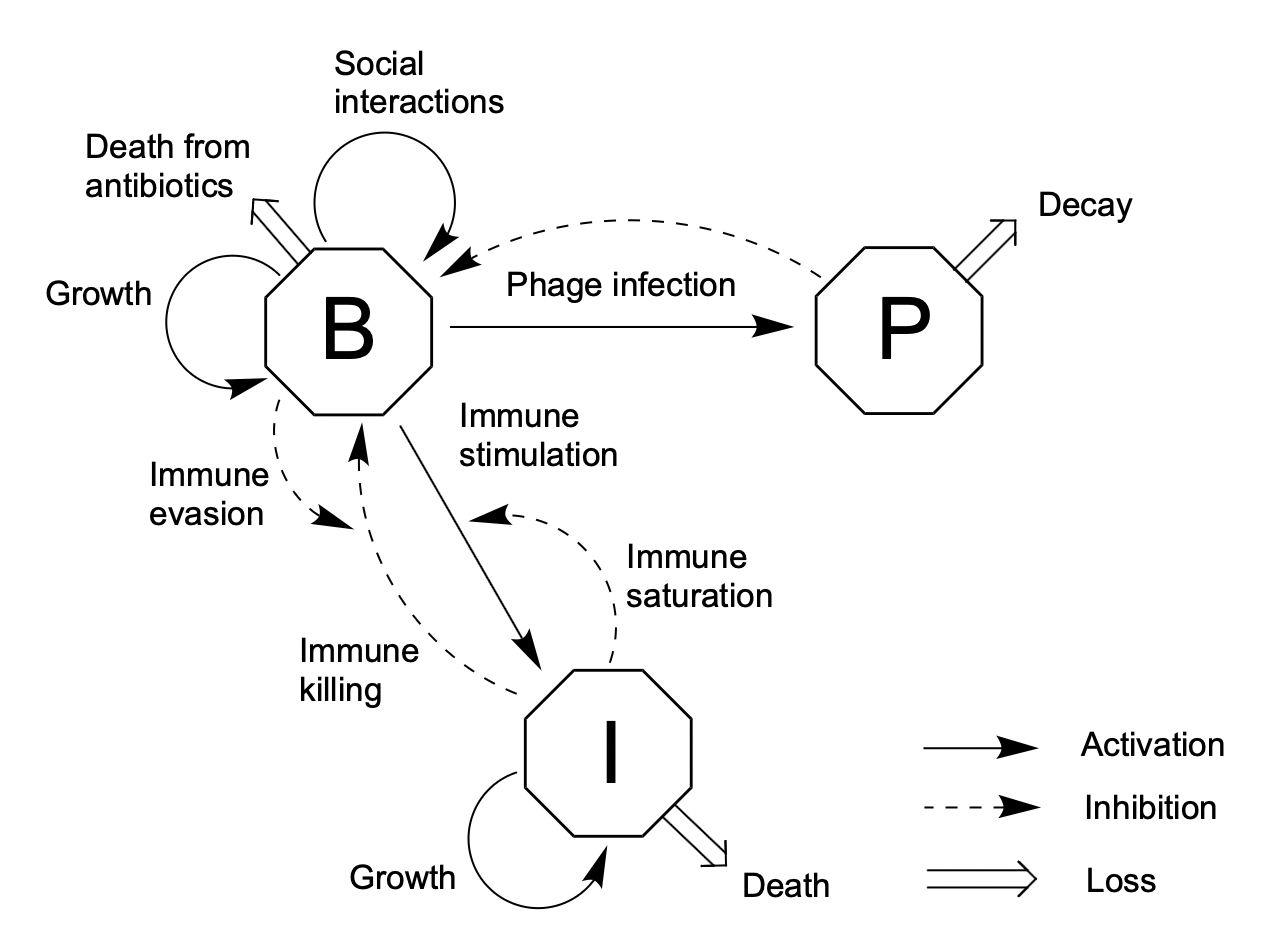}
    \caption{Diagram of the model. $B$, $I$, and $P$ denote bacteria, immune (effector) cells, and phages, respectively.}
    \label{fig:diagram}
\end{figure}

We designed a system of differential equations to model the dynamics between bacteria, a host's immune response, phages, and antibiotics, which is depicted graphically in Figure \ref{fig:diagram}. Bacteria are killed by immune system effector cells, phages, and antibiotics. However, they also undergo various social interactions among themselves that inhibit the immune response and phage predation (as detailed below). Effector cells are produced at a baseline rate by the host, but bacteria stimulate an immune response thereby increasing their production \citep{shepherd2020t}. Effector cells also die both naturally and by bacterial defenses such as toxins. Phages act as predators to bacterial prey in a Lotka-Volterra model, growing by killing bacteria as well as decaying. The variables $B$, $I$, and $P$ represent the concentrations (in millions per mL) of bacterial cells, immune (effector) cells, and phages, respectively, and the variable $A$ represents the concentration of antibiotics in units of mg per L.

The equations for the mathematical model were derived using considerations from previous models of immune system and pathogen dynamics in the literature \cite{leung2017modeling,kuznetsov1994nonlinear,shepherd2020t,barber2021predicting,reynolds2006reduced}. However, previous models of the dynamics of bacteria, the immune system, and phages assume immune saturation similar to a logistic equation \citep{leung2017modeling,rodriguez2020quantitative}. Furthermore, previous models have considered saturating phage adsorption rates using a Hill function of the phage density \citep{rodriguez2020quantitative,gomez2025mathematical}, which models the scenario where multiple phages may bind to the same bacterium thus reducing the effective infection rate. Here, we also assume a Hill function, but of bacterial density, which models the effects of protective social dynamics. Specifically, bacteria have been shown to reduce the number of receptors on their surfaces (phage entry points) in response to quorum sensing signals, halving the phage adsorption rate \citep{hoyland2017quorum}. Thus, we assume that the effective adsorption rate is $\phi/(1+B/\gamma)$ for parameters $\phi>0$ and $\gamma>0$.

The full set of equations for the mathematical model are:
\begin{subequations}
\begin{align}
    \dot{B} &= \overbrace{\rho B\left( 1 - \frac{B}{\kappa} \right)}^{\text{growth}} - \overbrace{\frac{\epsilon BI}{1+B/\zeta}}^{\substack{\text{death by}\\ \text{effector}}} - \overbrace{\frac{\phi BP}{1+B/\gamma}}^{\substack{\text{death by}\\ \text{phage}}} - \overbrace{\frac{\alpha BA^\tau}{A^\tau+\xi^\tau}}^{\substack{\text{death by}\\ \text{antibiotic}}}, \label{eq:dotB} \\
    \dot{I} &= I\bigg( \overbrace{\frac{\theta B}{B+\eta}}^{\text{activation}} - \overbrace{(\mu B+\delta)}^{\text{death}} \bigg) + \overbrace{\sigma,}^{\text{inflow}} \label{eq:dotI} \\
    \dot{P} &= \overbrace{\frac{\beta\phi BP}{1+B/\gamma}}^{\text{replication}} - \overbrace{\omega P,}^{\text{decay}} \label{eq:dotP} \\
    \dot{A} & = \overbrace{D(t)}^{\text{dosing}}-\overbrace{\nu A.}^{\text{decay}}. \label{eq:dotA}
\end{align}
\end{subequations}

In Equation \ref{eq:dotB}, the first term represents the growth rate of bacteria, which is modeled using a logistic growth function with a maximal growth rate of $\rho$ and carrying capacity $\kappa$. The remaining terms of this equation represent the rate bacteria are killed by effector cells and phages, which are reduced by large bacterial populations, and antibiotics. Equation \ref{eq:dotI} represents the net growth rate of the effector cells. The first term represents the activation of the immune response, the second term represents effector cell death, and the third term is the inflow of new effector cells produced by the host into the region. Finally, Equations \ref{eq:dotP} and \ref{eq:dotA} represent the net growth rate of phages as well as the dosing and decay of antibiotics, respectively. Antibiotics are assumed to be administered at regular intervals and are explored numerically in Section \ref{results:timeseries}. A summary of the parameters and their baseline values can be found in Table \ref{tbl:param}.

\begin{table}[ht!]
\centering
\begin{tabular}{clll}
\toprule
Parameter & Definition & Value & Units \\
\midrule
$\alpha$ & antibiotic kill rate & $0.1247$ & 1/hr \\ 
$\beta$ & phage burst size & $100$ & phages/cell \\ 
$\gamma$ & quorum sensing threshold & $100$ & $10^6$cells/mL \\ 
$\delta$ & effector death rate & $0.002$ & 1/hr \\ 
$\epsilon$ & effector kill rate & $0.082$ & (mL/$10^6$)/hr \\ 
$\zeta$ & immune half saturation & $2.2$ & $10^6$cells/mL \\ 
$\eta$ & activation parameter & $0.1$ & $10^6$cells/mL \\ 
$\theta$ & maximum activation rate & $0.97$ & 1/hr \\ 
$\kappa$ & carrying capacity of bacteria & $1000$ & $10^6$cells/mL \\ 
$\mu$ & inactivation rate & $0.01$ & (mL/$10^6$cells)/hr \\ 
$\nu$ & antibiotic decay rate & $0.34657$ & 1/hr \\ 
$\xi$ & antibiotic half saturation & $18.24$ & mg/L \\ 
$\rho$ & bacteria growth rate & $1$ & 1/hr \\ 
$\sigma$ & effector birth rate & $0.005$ & $10^6$cells/hr \\ 
$\tau$ & Hill coefficient & $1.416$ & --- \\ 
$\phi$ & phage adsorption rate & $0.05$ & 1/hr \\ 
$\omega$ & phage death rate & $1$ & 1/hr \\ 
\bottomrule
\end{tabular}
\caption{Summary definitions of parameters and variables.}\label{tbl:param}
\end{table}

The baseline parameter values for $\beta$, $\epsilon$, $\zeta$, $\eta$, $\theta$, $\kappa$, $\rho$, $\phi$, and $\omega$ are from the parameter values of the model in \cite{leung2017modeling} with data from other studies. Specifically, $\epsilon$ and $\zeta$ were adapted from \cite{drusano2011saturability}, $\eta$ was obtained from \cite{sadek1998chemokines} and \cite{wigginton2001model}, $\kappa$ was obtained from \cite{guo2011quantitative}, $\rho$ was obtained from \cite{guo2011quantitative}, $\phi$ was obtained from \cite{shao2008bacteriophage} and \cite{de2006viruses}, and $\omega$ was obtained from \cite{hodyra2015mammalian}. $\theta$ is taken from the maximum growth rate of the immune response calculated in \cite{leung2017modeling} using interstitial neutrophil recruitment data from \cite{reutershan05}. The immune system model of \cite{leung2017modeling} is similar to ours in that there is a Hill function for activation/stimulation of the immune system. However, their equation includes a density-dependent factor to model immune saturation, and it does not include innate or bacterial induced death rates, nor inflow of immune cells.

The remaining parameter values for the bacteria-immune system-phage system were determined as follows. The birth rate of effector cells $\sigma$ and the inactivation rate $\mu$ were taken from \cite{reynolds2006reduced}. The effector death rate $\delta$ was also taken from \cite{reynolds2006reduced}, sourced from \cite{Janeway2001} and \cite{Zouali2001}. It is important to note that the immune system is modeled phenomenologically in this model instead of being cell-specific. Quorum sensing activation is typically observed at bacterial cell densities between $10^7$ and $10^9$ cells/mL \citep{xayarath2016being}. We assume it occurs on the order of $10^8$ bacteria/mL for our baseline scenario, although we explore both an order of magnitude lower and higher in our simulations. Since the effective adsorption rate halved at this value, $\gamma=100$.

Finally, the antibiotic kill rate, half saturation rate, and Hill coefficient for the antibiotic cefoxitin were obtained from \cite{ferro2015time}. Cefoxitin was chosen as the antibiotic to be modeled in the simulations due to it being fairly commonly used in clinical settings as well as the fact that there are many bacteria that are resistant to it \citep{tebano2024antibiotic,sartelli2023six}. The antibiotic decay rate $\nu$ comes from \cite{brouwers2020stability}. Antibiotics are assumed to be administered in doses of $100$ mg/L, and the initial dose is given at $100$ hours. This dosage delay was selected because a study reviewing the median time from bacterial infection to antibiotic treatment found that on average antibiotic treatment is administered approximately $100$ hours after initial exposure to the bacterial infection \citep{holty2006systematic}. After this initial dose, recurring doses are administered on a schedule every $24$ hours to simulate an antibiotic that is taken once a day. Similarly, we also assume that phages are introduced as a treatment at $100$ hours, though only once as they replicate thereafter.

\section{Results}

\subsection{Equilibria and stability analysis} \label{results:analytical}

Here, we analyze the system without the presence of antibiotics. Equilibria occur when:
\begin{subequations}
\begin{align}
    0 &= B\left(\rho\left( 1 - \frac{B}{\kappa} \right) - \frac{\epsilon I}{1+B/\zeta} - \frac{\phi P}{1+B/\gamma}\right), \\
    0 &= I\left( \frac{\theta B}{B+\eta} - \mu B -\delta \right) + \sigma, \\
    0 &= P\left(\frac{\beta\phi B}{1+B/\gamma} - \omega\right).
\end{align}
\end{subequations}
We begin by discussing the bacteria-free and phage-free equilibrium $\mathcal{E}_1 = (\bar{B}_1,\bar{I}_1,\bar{P}_1) = (0,\sigma/\delta,0)$. Following this we will discuss the phage-free bacterial infection equilibria ($\mathcal{E}_2 = (\bar{B}_2,\bar{I}_2,0)$), and the coexistence of bacteria, effectors, and phages ($\mathcal{E}_3 = (\bar{B}_3,\bar{I}_3,\bar{P}_3)$). $\mathcal{E}_1$ and $\mathcal{E}_3$, if they exist, are unique, while there may be one or three equilibria that satisfy $\mathcal{E}_2$. We label these equilibria $\mathcal{E}_{2a}$, $\mathcal{E}_{2b}$, and $\mathcal{E}_{2c}$. We also note that the Jacobian matrix is:
\begin{equation}
    J = \begin{bmatrix}
        \rho\left(1 - \dfrac{2B}{\kappa}\right) - \dfrac{\epsilon I}{(1+B/\zeta)^2} - \dfrac{\phi P}{(1+B/\gamma)^2} & - \dfrac{\epsilon B}{1+B/\zeta} & - \dfrac{\phi B}{1+B/\gamma} \\
        \left(\dfrac{\theta \eta}{(B+\eta)^2} - \mu\right)I & \dfrac{\theta B}{B+\eta} - \mu B -\delta & 0 \\
        \dfrac{\beta\phi P}{(1+B/\gamma)^2} & 0 & \dfrac{\beta\phi B}{1+B/\gamma} - \omega
    \end{bmatrix}.
\end{equation}

Evaluating the Jacobian at $\mathcal{E}_1$ gives us:
\begin{equation}
    J(\mathcal{E}_1) = \begin{bmatrix}
        \rho - \epsilon\bar{I}_1 & 0 & 0 \\
        \left(\dfrac{\theta}{\eta} - \mu\right)\bar{I}_1 & -\delta & 0 \\
        0 & 0 & - \omega
    \end{bmatrix}.
\end{equation}
The characteristic equation for this matrix is:
\begin{equation}
    \left(\lambda - \rho + \epsilon\bar{I}_1 \right)(\lambda + \delta)(\lambda + \omega) = 0.
\end{equation}
This has two negative eigenvalues ($\lambda_2 = -\delta, \lambda_3 = -\omega$) and the last is positive if $\rho > \epsilon\bar{I}_1$. Therefore, since we are concerned with the case where bacteria are able to colonize the body, we will assume this inequality holds true for the remainder of the analysis.

\begin{table}[ht!]
\centering
\begin{tabular}{cccccc}
\toprule
$\bar{B}_2^4$ term & $\bar{B}_2^3$ term & $\bar{B}_2^2$ term & $\bar{B}_2$ term & constant term & sign changes \\
\midrule
$+$ & $+$ & $+$ & $+$ & $-$ & $1$ \\
$+$ & $+$ & $+$ & $-$ & $-$ & $1$ \\
$+$ & $+$ & $-$ & $+$ & $-$ & $3$ \\
$+$ & $+$ & $-$ & $-$ & $-$ & $1$ \\
$+$ & $-$ & $+$ & $+$ & $-$ & $3$ \\
$+$ & $-$ & $+$ & $-$ & $-$ & $3$ \\
$+$ & $-$ & $-$ & $+$ & $-$ & $3$ \\
$+$ & $-$ & $-$ & $-$ & $-$ & $1$ \\
\bottomrule
\end{tabular}
\caption{Sign and sign changes of the coefficients of Equation \ref{eq:polynomial}.}\label{tbl:signs}
\end{table}

Next, consider $\mathcal{E}_2 = (\bar{B}_2,\bar{I}_2,0)$. Solving $\dot{B}=0$, we have $\bar{I}_2 = \rho(1-\bar{B}_2/\kappa)(1+\bar{B}_2/\zeta)/\epsilon$. Plugging this into $\dot{I}=0$ and simplifying, we get:
\begin{multline}
    \frac{\mu}{\zeta\kappa}\bar{B}_2^4 + \left( \frac{\eta\mu+\delta-\theta}{\zeta\kappa} + \frac{\mu}{\kappa} - \frac{\mu}{\zeta} \right)\bar{B}_2^3 + \left( \frac{\delta\eta}{\zeta\kappa} + (\theta-\eta\mu-\delta)\left(\frac{1}{\zeta}-\frac{1}{\kappa}\right) - \mu \right)\bar{B}_2^2 \\
    + \left( \theta-\eta\mu-\delta - \delta\eta\left(\frac{1}{\zeta} - \frac{1}{\kappa}\right) + \frac{\epsilon\sigma}{\rho} \right)\bar{B}_2 + \frac{\delta\eta}{\rho}\left(\epsilon\bar{I}_1-\rho\right) = 0. \label{eq:polynomial}
\end{multline}
Note that the coefficient of the first term in the polynomial is positive and the last is negative assuming $\rho > \epsilon\bar{I}_1$. Thus, by Descartes' Rule of Signs, there are either one or three positive real roots as shown in Table \ref{tbl:signs}. Many of the signs of these terms do not have simple and specific biological meaning. Nonetheless, we can make some general observations. For one, the coefficient of $\bar{B}_2^3$ is negative when activation of the immune system is sufficiently rapid, which occurs for high $\theta$ and low $\eta$. The effects of high immune activation, however, on coefficients for $\bar{B}_2^2$ and $\bar{B}_2$ are more complicated, either increasing or decreasing them. Although, for the baseline parameters in Table \ref{tbl:param}, $\zeta < \kappa$ and thus high activation increases the coefficient of $\bar{B}_2$.

Returning to our stability analysis, the Jacobian evaluated at $\mathcal{E}_2$ is:
\begin{equation}
    J(\mathcal{E}_2) = \begin{bmatrix}
        \rho\left(1 - \dfrac{2\bar{B}_2}{\kappa}\right) - \dfrac{\epsilon\bar{I}_2}{(1+\bar{B}_2/\zeta)^2} & - \dfrac{\epsilon\bar{B}_2}{1+\bar{B}_2/\zeta} & - \dfrac{\phi\bar{B}_2}{1+\bar{B}_2/\gamma} \\
        \left(\dfrac{\theta\eta}{(\bar{B}_2+\eta)^2} - \mu\right)\bar{I}_2 & -\dfrac{\sigma}{\bar{I}_2} & 0 \\
        0 & 0 & \dfrac{\beta\phi\bar{B}_2}{1+\bar{B}_2/\gamma} - \omega
    \end{bmatrix}.
\end{equation}
And the characteristic equation is:
\begin{multline}
    \left( \lambda - \frac{\beta\phi\bar{B}_2}{1+\bar{B}_2/\gamma} + \omega \right)\left( \left( \lambda - \rho\left(1 - \dfrac{2\bar{B}_2}{\kappa}\right) + \dfrac{\epsilon\bar{I}_2}{(1+\bar{B}_2/\zeta)^2} \right)\left( \lambda + \frac{\sigma}{\bar{I}_2} \right)\right. \\
    \left. + \left(\dfrac{\theta\eta}{(\bar{B}_2+\eta)^2} - \mu\right)\dfrac{\epsilon\bar{B}_2\bar{I}_2}{1+\bar{B}_2/\zeta} \right) = 0.
\end{multline}
From this we can see that $\bar{B}_2 > 1/(\beta\phi/\omega-1/\gamma)$ is a sufficient condition for instability of $\mathcal{E}_2$. We numerically explore the eigenvalues further in the next section.

Now, consider the case where there is coexistence of all three types ($\mathcal{E}_3 = (\bar{B}_3,\bar{I}_3,\bar{P}_3)$). Solving for $\dot{P}=0$ gives us $\bar{B}_3=1/(\beta\phi/\omega-1/\gamma)$. Then, we have:
\begin{subequations}
\begin{align}
    \bar{I}_3 &= \frac{\sigma}{\mu\bar{B}_3 + \delta - \theta \bar{B}_3/(\bar{B}_3+\eta)}, \\
    \bar{P}_3 &= \frac{1+\bar{B}_3/\gamma}{\phi}\left(\rho\left( 1 - \frac{\bar{B}_3}{\kappa} \right) - \frac{\epsilon\bar{I}_3}{1+\bar{B}_3/\zeta}\right),
\end{align}
\end{subequations}
assuming that these are positive.

The Jacobian at this equilibrium is:
\begin{equation}
    J(\mathcal{E}_3) = \begin{bmatrix}
        \rho\left(1 - \dfrac{2\bar{B}_3}{\kappa}\right) - \dfrac{\epsilon\bar{I}_3}{(1+\bar{B}_3/\zeta)^2} - \dfrac{\phi \bar{P}_3}{(1+\bar{B}_3/\gamma)^2} & - \dfrac{\epsilon\bar{B}_3}{1+\bar{B}_3/\zeta} & - \dfrac{\phi\bar{B}_3}{1+\bar{B}_3/\gamma} \\
        \left(\dfrac{\theta\eta}{(\bar{B}_3+\eta)^2} - \mu\right)\bar{I}_3 & -\dfrac{\sigma}{\bar{I}_3} & 0 \\
        \dfrac{\beta\phi \bar{P}_3}{(1+\bar{B}_3/\gamma)^2}  & 0 & 0
    \end{bmatrix},
\end{equation}
and it has the characteristic equation:
\begin{equation}
    \left( \lambda + \dfrac{\sigma}{\bar{I}_3} \right)\left( \lambda^2 + \left( \dfrac{\epsilon\bar{I}_3}{(1+\bar{B}_3/\zeta)^2} + \dfrac{\phi\bar{P}_3}{(1+\bar{B}_3/\gamma)^2} - \rho\left(1 - \dfrac{2\bar{B}_3}{\kappa}\right) \right)\lambda + \frac{\beta\phi^2\bar{B}_3\bar{P}_3}{(1+\bar{B}_3/\gamma)^3}\right) = 0. \label{eq:char_eq_E3}
\end{equation}
Note that $\lambda_1 = -\sigma/\bar{I}_3 < 0$ and:
\begin{equation}
    \dfrac{\epsilon\bar{I}_3}{(1+\bar{B}_3/\zeta)^2} + \dfrac{\phi\bar{P}_3}{(1+\bar{B}_3/\gamma)^2} - \rho\left(1 - \dfrac{2\bar{B}_3}{\kappa}\right) =\bar{B}_3\left(-\dfrac{\epsilon\bar{I}_3/\zeta}{(1+\bar{B}_3/\zeta)^2} - \dfrac{\phi\bar{P}_3/\gamma}{(1+\bar{B}_3/\gamma)^2} + \frac{\rho}{\kappa}\right).
\end{equation}
If $\gamma,\zeta \gg 0$, then all the coefficients of the quadratic in Equation \ref{eq:char_eq_E3} are positive, and thus the equilibrium is stable. Otherwise, it may not be. We explore the eigenvalues numerically in the next section.

\subsection{Numerical results} \label{results:numerical}

\subsubsection{Bifurcation diagrams}

\begin{figure}[ht!]
\captionsetup[subfigure]{justification=centering}
    \centering
    \begin{subfigure}[]{0.32\columnwidth}
        \includegraphics[width=\textwidth]{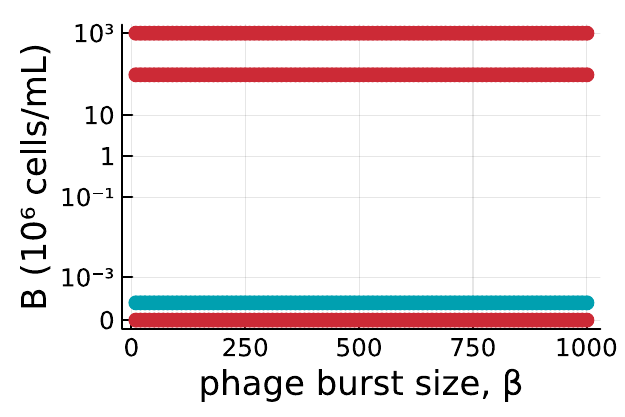}
    \end{subfigure}
    \begin{subfigure}[]{0.32\columnwidth}
        \includegraphics[width=\textwidth]{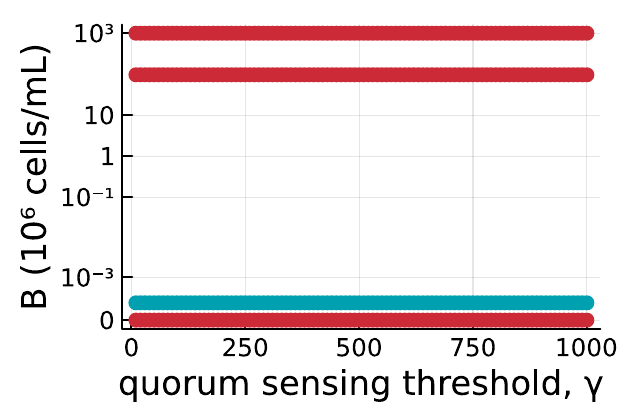}
    \end{subfigure}
        \begin{subfigure}[]{0.32\columnwidth}
        \includegraphics[width=\textwidth]{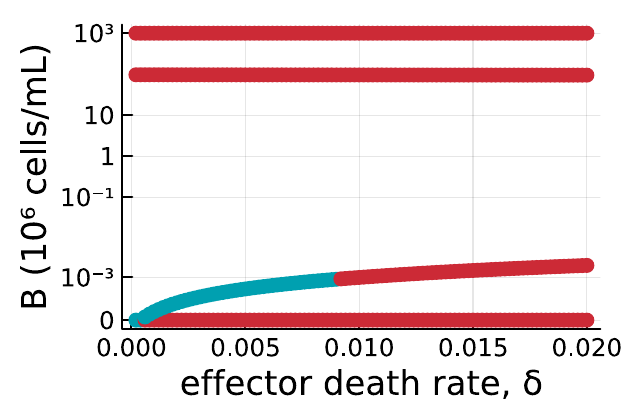}
    \end{subfigure} \\
    \begin{subfigure}[]{0.32\columnwidth}
        \includegraphics[width=\textwidth]{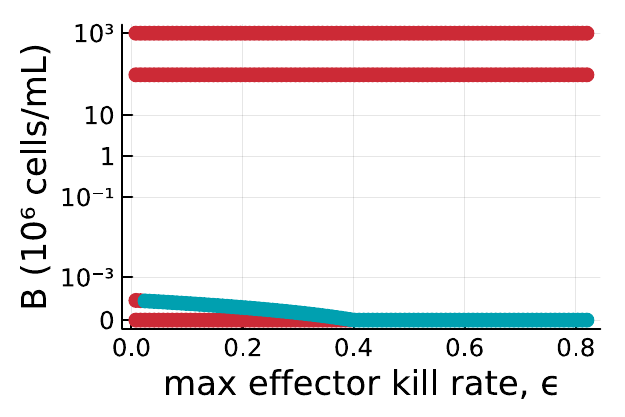}
    \end{subfigure}
    \begin{subfigure}[]{0.32\columnwidth}
        \includegraphics[width=\textwidth]{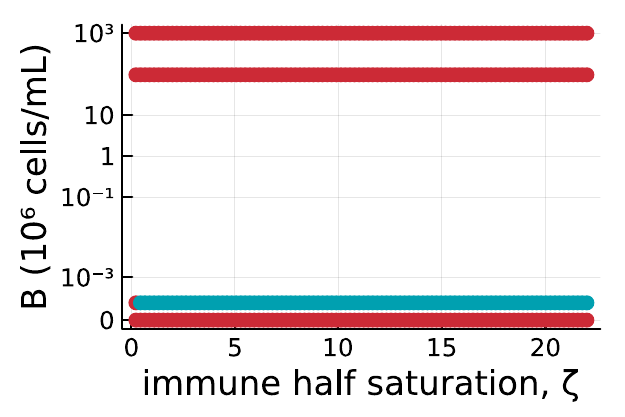}
    \end{subfigure}
        \begin{subfigure}[]{0.32\columnwidth}
        \includegraphics[width=\textwidth]{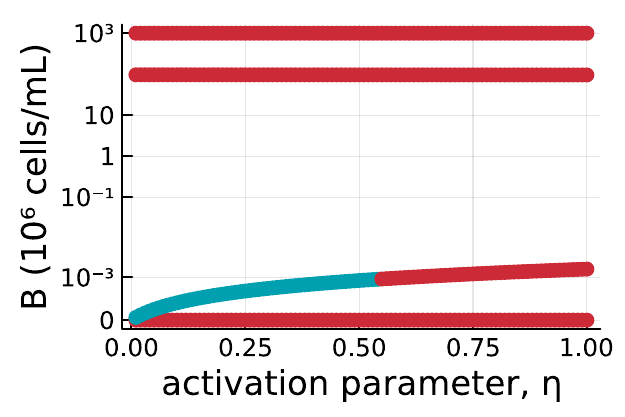}
    \end{subfigure} \\
    \begin{subfigure}[]{0.32\columnwidth}
        \includegraphics[width=\textwidth]{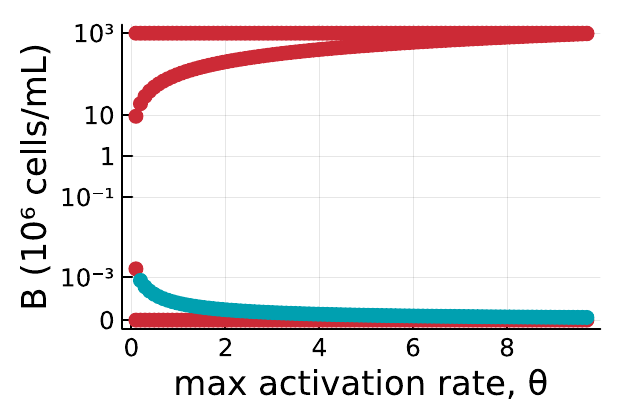}
    \end{subfigure}
    \begin{subfigure}[]{0.32\columnwidth}
        \includegraphics[width=\textwidth]{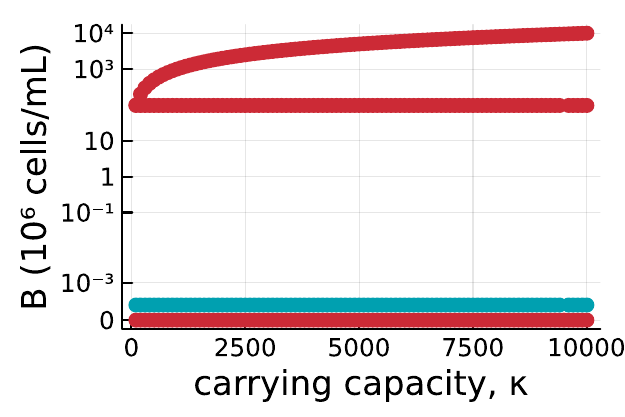}
    \end{subfigure}
        \begin{subfigure}[]{0.32\columnwidth}
        \includegraphics[width=\textwidth]{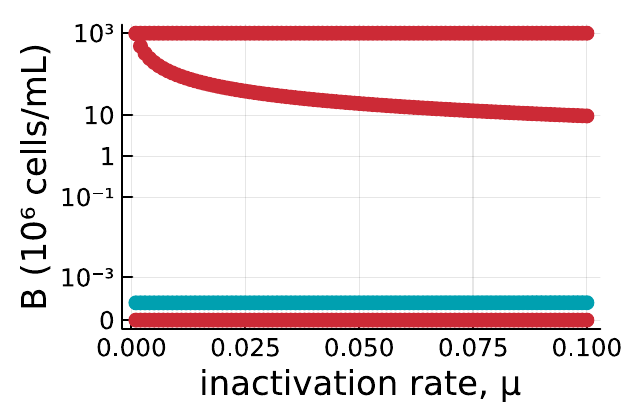}
    \end{subfigure} \\
    \begin{subfigure}[]{0.32\columnwidth}
        \includegraphics[width=\textwidth]{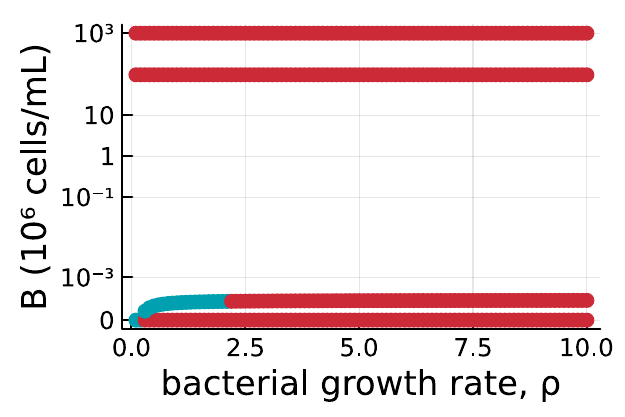}
    \end{subfigure}
    \begin{subfigure}[]{0.32\columnwidth}
        \includegraphics[width=\textwidth]{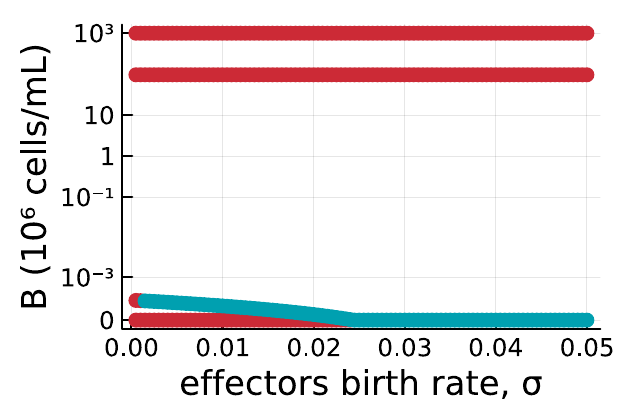}
    \end{subfigure}
    \begin{subfigure}[]{0.32\columnwidth}
        \includegraphics[width=\textwidth]{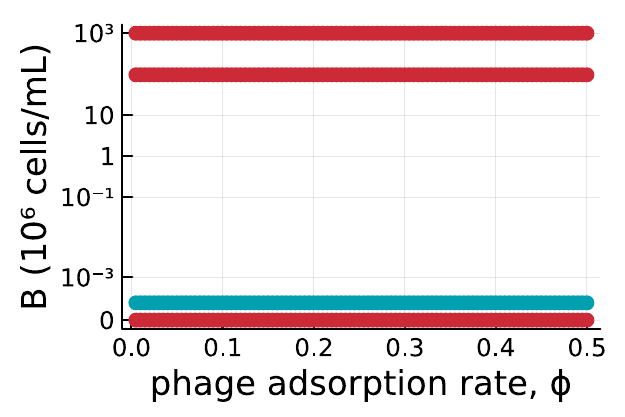}
    \end{subfigure} \\
    \begin{subfigure}[]{0.32\columnwidth}
        \includegraphics[width=\textwidth]{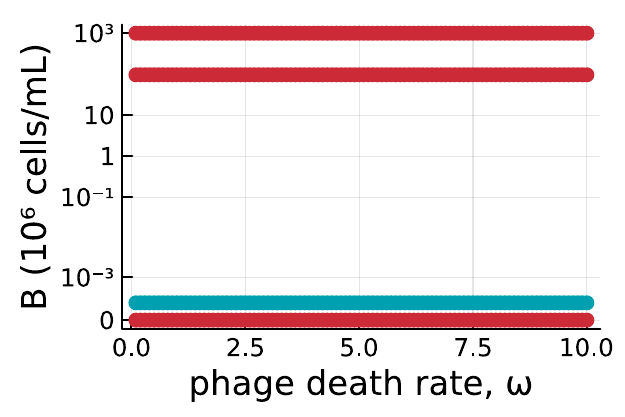}
    \end{subfigure} 
    \caption{Bifurcation diagrams for equilibrium number of bacteria $\bar{B}_i$. Parameters are varied from one tenth to ten times the baseline values in Table \ref{tbl:param}. Blue indicates a stable equilbrium and red a saddle. The vertical axis is plotted in symmetrical log scale: data is converted via the function $\text{symlog}(x) = \log_{10}(1+10^3x)$.}
    \label{fig:bifurcation}
\end{figure}

We begin with an exploration of parameter space by individually varying the parameters from Table \ref{tbl:param} before considering specific scenarios. Figure \ref{fig:bifurcation} depicts these bifurcation diagrams. Red corresponds to unstable equilibria and blue to stable equilibria. This results in either a low bacterial population density or even no bacteria altogether. It is important to note that the y-axes of these plots correspond with the bacterial density measured in millions of cells per mL. For example, a bacterial density of $10^{-3}$ indicates a bacterial density of $10^3$ cells per mL.

Across all diagrams, there is either a single stable equilibrium with a relatively low density of bacteria or no stable equilibria. The latter scenario results in oscillations, which we will discuss in depth in Section \ref{results:timeseries}. The bifurcation plots for $\beta$, $\gamma$, $\kappa$, $\mu$, $\phi$, and $\omega$ all lack a bifurcation point where the stable equilibrium becomes unstable, showing a robustness of the qualitative results with respect to these parameters. The remaining parameters exhibit bifurcation points. For $\delta$, we observe a bifurcation point at $\delta \approx 0.009$. Below this point, there is the single stable equilibrium and above it all equilibria are unstable. $\epsilon$ has a bifurcation point corresponding with low densities of bacteria at $\epsilon \approx 0.01$. Likewise, the bifurcation plot of $\zeta$ shows a bifurcation point at $\zeta \approx 0.1$. The bifurcation plot of $\eta$ indicates a bifurcation point corresponding to a bacterial density of $10^{-3}$ and $\eta \approx 0.55$. At very low values of $\theta$ and a bacterial density of approximately $10^{-3}$, a bifurcation point exists for the parameter $\theta$. For the parameter $\rho$, there appear to be two bifurcation points. One of these points corresponds with the bacteria free state and very small values of $\rho$. This result mirrors our analytical findings: for sufficiently low $\rho$, the bacteria-free equilibrium is the sole stable equilibrium. The other bifurcation point occurs at low bacterial densities and when $\rho \approx 2$. Finally, the bifurcation diagram generated for $\sigma$ shows a bifurcation point at low bacterial densities and very small values of $\sigma$.

\subsubsection{Time series} \label{results:timeseries}

\begin{figure}[ht!]
\captionsetup[subfigure]{justification=centering}
    \centering
    \begin{subfigure}[]{0.4\columnwidth}
        \caption{$B$ vs time}\label{fig:timeseries_B_antibiotics_only}
        \includegraphics[width=\textwidth]{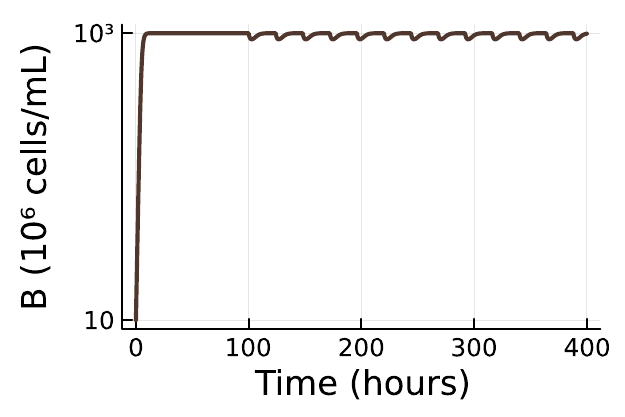}
    \end{subfigure}
    \begin{subfigure}[]{0.4\columnwidth}
        \caption{$I$ vs time}\label{fig:timeseries_I_antibiotics_only}
        \includegraphics[width=\textwidth]{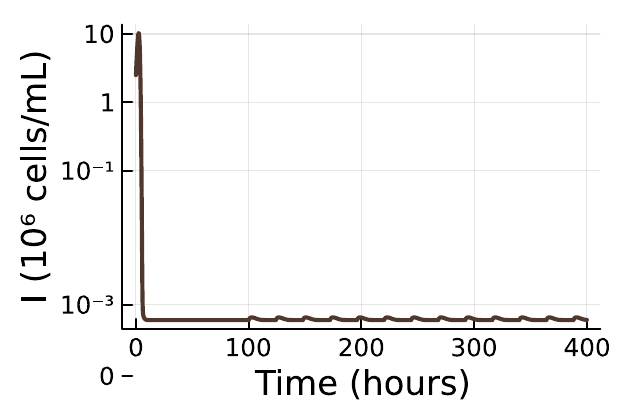}
    \end{subfigure} \\
        \begin{subfigure}[]{0.4\columnwidth}
        \caption{$A$ vs time}\label{fig:timeseries_A_antibiotics_only}
        \includegraphics[width=\textwidth]{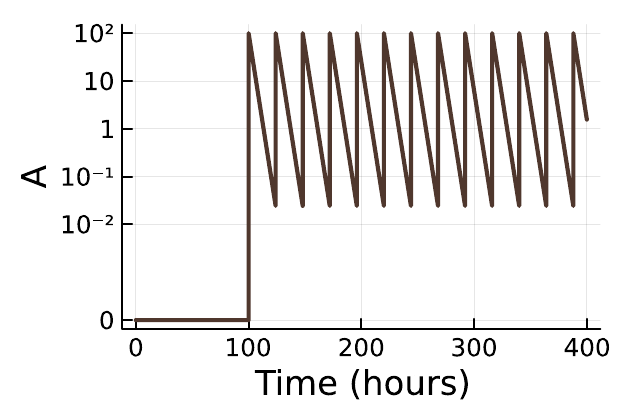}
    \end{subfigure}
    \caption{Time series plots for the case with periodic applications of antibiotics, but no phages are present. Parameters are taken from Table \ref{tbl:param}. Antibiotics alone are unable to control and suppress the infection. The vertical axis is plotted in symmetrical log scale: data is converted via the function $\text{symlog}(x) = \log_{10}(1+10^3x)$.}
    \label{fig:timeseries_antibiotics_only}
\end{figure}

In this section, we begin with a demonstration of the case where no phages are present and only antibiotics are used to attempt to suppress the infection. Figure \ref{fig:timeseries_antibiotics_only} depicts the time series plots for this case. The initial conditions are $B_0=10$ and $I_0 = \sigma/\delta$, which allow for the bacteria to escape immune suppression, approach the carrying capacity, and enervate the immune response. After $100$ hours, the antibiotic is applied at regular daily intervals. However, this therapy is insufficient in quelling the infection.

\begin{figure}[ht!]
\captionsetup[subfigure]{justification=centering}
    \centering
    \begin{subfigure}[]{0.4\columnwidth}
        \caption{$B$ vs time}\label{fig:timeseries_B_baseline}
        \includegraphics[width=\textwidth]{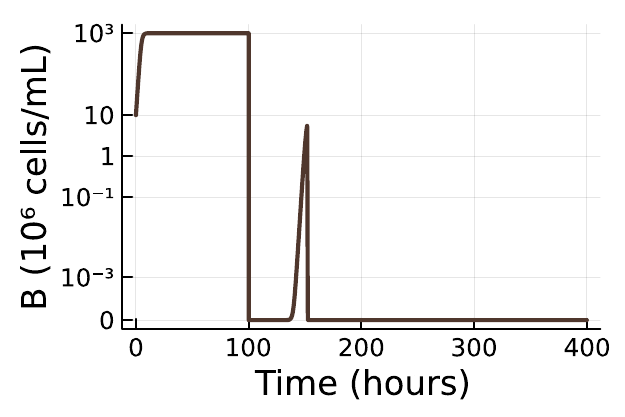}
    \end{subfigure}
    \begin{subfigure}[]{0.4\columnwidth}
        \caption{$I$ vs time}\label{fig:timeseries_I_baseline}
        \includegraphics[width=\textwidth]{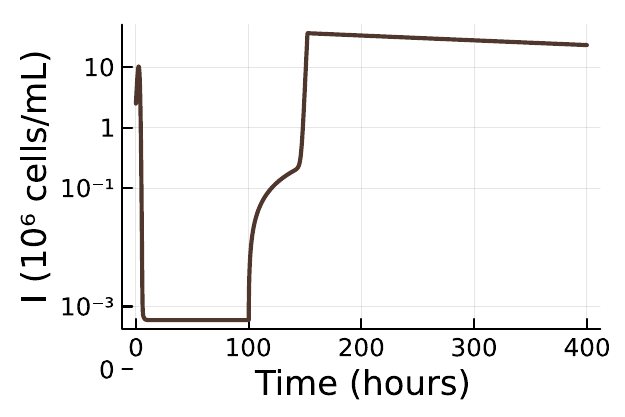}
    \end{subfigure} \\
        \begin{subfigure}[]{0.4\columnwidth}
        \caption{$P$ vs time}\label{fig:timeseries_P_baseline}
        \includegraphics[width=\textwidth]{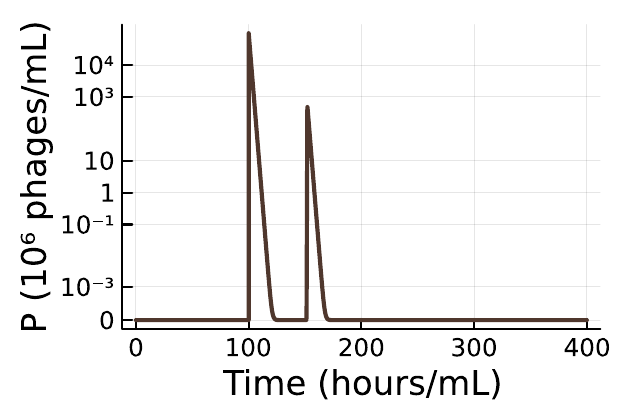}
    \end{subfigure}
    \caption{Time series plots for the baseline scenario with parameters from Table \ref{tbl:param}. Phage therapy successfully assists the immune system in suppressing the infection. The vertical axis is plotted in symmetrical log scale: data is converted via the function $\text{symlog}(x) = \log_{10}(1+10^3x)$.}
    \label{fig:timeseries_baseline}
\end{figure}

\begin{table}[ht!]
\centering
\begin{tabular}{ccccl}
\toprule
Equilibrium & B & I & P & Eigenvalues \\
\midrule
$\mathcal{E}_1$ & $0$ & $2.5$ & $0$ & 2 negative real and 1 positive real  \\
$\mathcal{E}_{2a}$ & $0.0001643598$ & $12.19603$ & $0$ & \begin{tabular}{@{}l@{}}1 negative real, complex pair\\ with negative real part\end{tabular} \\
$\mathcal{E}_{2b}$ & $96.70080$ & $495.21628$ & $0$ & 2 positive real, 1 negative real \\
$\mathcal{E}_{2c}$ & $999.9999$ & $0.0005535814$ & $0$ & 1 positive real, 2 negative real \\
$\mathcal{E}_3$ & DNE & DNE & DNE & N/A \\
\bottomrule
\end{tabular}
\caption{Summary of equilibria found numerically.}\label{tbl:equilibria_baseline}
\end{table}

Next, consider a phage treatment with the same initial conditions of $B_0=10$ and $I_0=\sigma/\delta$ where treatment is initiated after $100$ hours. As seen in the time series plots in Figure \ref{fig:timeseries_baseline}, phage therapy is able to control the infection. Phage density increases rapidly, bacterial density crashes, and effector cell density rebounds. Indeed, with the presence of phages, the system has a sole stable equilibrium with low bacterial density. Table \ref{tbl:equilibria_baseline} details all of the numerically found equilibria, of which there are four, and the nature of their eigenvalues. The first equilibrium ($\mathcal{E}_1$) is the bacteria and phage free state, which is unstable. Two of the three phage-free equilibria ($\mathcal{E}_{2b}$ and $\mathcal{E}_{2c}$) are saddle points, and the other equilibrium ($\mathcal{E}_{2a}$) is stable. Phages are unable to persist with bacteria at this density. Nonetheless, phages are crucial in undermining the stability of the high bacterial density equilibrium ($\mathcal{E}_{2c}$). The last potential equilibrium point, corresponding to the coexistence of bacteria, effectors, and phages, does not exist for these parameter values ($\bar{I}_3$ is negative). Although bacteria do still persist at low levels in this scenario, there are practical considerations that suggest they would be eliminated. In examining the time series plots, the minimum bacterial density reaches approximately $10^{-18}$ cells/mL after the introduction of phages. Though it oscillates as it equilibrates to approximately $164$ bacteria per mL, the bacteria would in effect be eliminated earlier.

\begin{figure}[ht!]
\captionsetup[subfigure]{justification=centering}
    \centering
    \begin{subfigure}[]{0.4\columnwidth}
        \caption{$B$ vs time}\label{fig:timeseries_B_immunodef}
        \includegraphics[width=\textwidth]{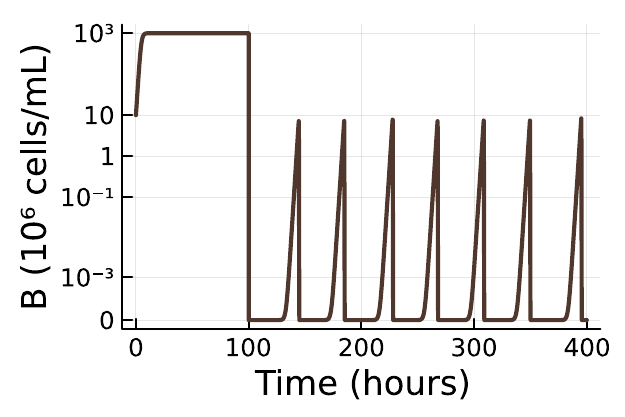}
    \end{subfigure}
    \begin{subfigure}[]{0.4\columnwidth}
        \caption{$I$ vs time}\label{fig:timeseries_I_immunodef}
        \includegraphics[width=\textwidth]{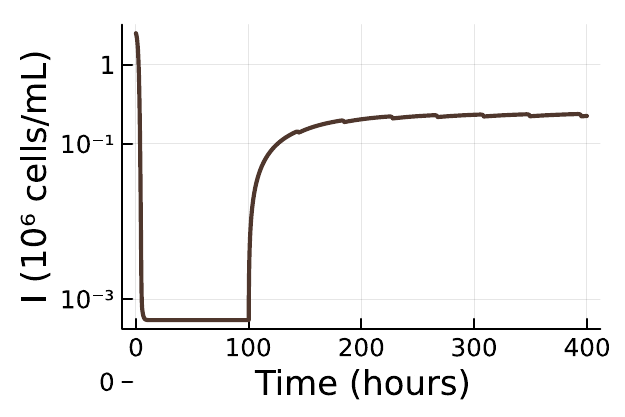}
    \end{subfigure} \\
        \begin{subfigure}[]{0.4\columnwidth}
        \caption{$P$ vs time}\label{fig:timeseries_P_immunodef}
        \includegraphics[width=\textwidth]{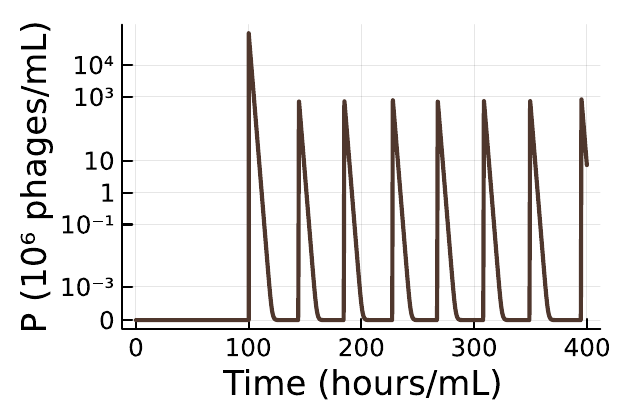}
    \end{subfigure}
    \caption{Time series plots for the immunodeficient host scenario. $\delta=0.02$, $\eta=1.0$, and $\theta=0.01$ while remaining parameters are taken from Table \ref{tbl:param}. The system is oscillatory. The vertical axis is plotted in symmetrical log scale: data is converted via the function $\text{symlog}(x) = \log_{10}(1+10^3x)$.}
    \label{fig:timeseries_immunodef}
\end{figure}

The results above reflect the scenarios from the bifurcation diagrams with only a single stable equilibrium at relatively low bacterial densities. We now consider time series for the second regime where there are no stable equilibria (using the same initial conditions and timing of treatment). Note that a few of these correspond to immunodeficient hosts: effector activation is low, which can occur via low $\theta$ or high $\eta$, or effector death rates are high. High bacterial growth rates can also result in no stable equilibria. Under these scenarios, the population density oscillates. Figure 
\ref{fig:timeseries_B_immunodef} depicts this scenario with $\delta=0.02$, $\eta=1.0$, and $\theta=0.01$. These oscillations bring bacterial density to elimination levels (approximately $10^{-13}$ bacteria per mL). A theoretical novelty of the system is that the results suggest chaotic behaviour. We numerically calculated the Lyapunov exponent, finding that it was positive, indicating chaos. However, as stated above, population levels reach extremely low values and thus the bacteria are effectively eliminated. A combination therapy of phages and antibiotics can drive bacterial densities even lower (the results for this case and apparent chaos are not shown here, although they are reproducible via the code).

Interestingly, these large oscillations are in large part due to quorum sensing mediated bacterial protections from phages. For comparison, consider the model of \cite{leung2017modeling}, which features a Holling type I functional response. It exhibits instability of oscillations when there are only bacteria and phages interacting. However, this occurs on long time scales. The Holling type II functional response that we consider accelerates this instability. By weakening phage predation, bacteria can grow to higher densities before being subdued by phages, which in turn results in large phage populations. Bacteria then lose their defenses as their density drops below the quorum sensing threshold, exacerbating their decline. The short term benefits of phage protection ultimately dooms the bacteria to elimination. Furthermore, there is a game theory insight into this result. In a heterogeneous population of bacteria, there is an incentive for individual bacterial clusters to quorum sense for their own immediate protection, even though abstaining would benefit the population as a whole. In a sense, bacteria that reduce their receptors due to quorum-sensing are free-riders or defectors, while those that do not are cooperators.

\section{Discussion}

The dynamics behind phage therapy in clinical settings are not fully understood, since the exact interactions between bacteria, immune cells, and phages are oftentimes very complex. To better understand these dynamics, we developed a model that incorporates the social behaviors of bacteria, an innate immune response, phages, and periodic antibiotic dosing. This model is similar to \cite{leung2017modeling}, though with different assumptions regarding the immune response. We assume a constant inflow of effector cells into the infected region along with bacteria-mediated death of effector cells. Structurally, our model, not including the phages, is similar to the immuno-tumor model of \cite{kuznetsov1994nonlinear}, which has also been extended to explore social dynamics \citep{morsky2018cheater}.

Our results indicate that the system has a single stable equilibrium for our baseline parameters in which effectors and low levels of bacteria coexist. Phage therapy has effectively eliminated the stable equilibrium with a large population of bacteria, and the immune system is thus able to contain the bacterial infection. Even for immunodeficient hosts, phage therapy results in such large oscillations in the population of bacteria that they would effectively be eliminated. This effect is further aided by the assistance of antibiotics in a combination therapy. This result has similarity to the finding from \cite{leung2017modeling} where it was observed that phage therapy and the host's immune system were able to eradicate the bacterial infection in the normal mice population but not in the neutropenic mice population. However, the presence of quorum-sensing mediated bacterial defenses in our model essentially results in elimination of the infection even in immunodeficient hosts.

The proposed model allows for analytical and numerical solutions to be found for the conditions where phages and the immune response can kill bacteria. However, this synergistic effect is a simplification made to be able to analyze a much more complex dynamic. One assumption that was made is that there is only one type of bacteria in the system that has a consistent reaction to the phages and immune response. In reality, the interactions between bacteria, an immune system, and phages are much more complex. In particular, bacteria are able to develop phage resistance by modifying the cell surface receptors that phages attach onto \citep{labrie2010bacteriophage}. However, it has been noted that increased phage resistance oftentimes leads to increased susceptibility to antibiotics \citep{chan2016phage}. As a result, a future study could explore an expansion of the model presented here with multiple bacterial species, mutations, and trade-offs for resistance to phages and antibiotics. Combination therapies that leverage these trade-offs --- such as switching environmental regimes \citep{morsky2022suppressing} --- could be explored to determine optimal treatments to control the bacteria. Finally, it has been demonstrated that both the innate and adaptive immune responses can actively eliminate phages \citep{hodyra2015mammalian}. This process can and would likely impact the overall effectiveness of phage therapy in a clinical context. In our model, the death rate of phages is assumed to be constant, represented by the parameter $\omega$, and determined by the density of bacteria. However, phages could also be killed by the immune system and help activate it. In this case, the effectiveness of phage therapy could be tied to the relationship between the killing of phages by the immune system and their synergistic elimination of bacteria.

\subsection*{Code and Data Availability}
\raggedright
Numerical methods in Julia were used to identify equilibria and characterize their stability. Specifically, the DifferentialEquations.jl package \citep{rackauckas2017differentialequations} was employed to numerically solve the system. The code to solve the equations numerically and produce the plots is available at github.com/bmorsky/phage-therapy.

\bibliographystyle{apalike}
\bibliography{microbes}

\end{document}